# Quantum interference in superposed lattices

Yejun Feng[1,*], Yishu Wang[2,3], T. F. Rosenbaum[4], P. B. Littlewood[5], Hua Chen[6]

[1]Okinawa Institute of Science and Technology Graduate University, Onna, Okinawa 904-0495, Japan
[2]Department of Materials Science and Engineering, University of Tennessee, Knoxville, Tennessee 37996, USA
[3]Department of Physics and Astronomy, University of Tennessee, Knoxville, Tennessee 37996, USA
[4]Division of Physics, Mathematics, and Astronomy, California Institute of Technology, Pasadena, California 91125, USA
[5]The James Franck Institute and Department of Physics, The University of Chicago, Illinois 60637, USA
[6]Department of Physics, Colorado State University, Fort Collins, Colorado 80523, USA
*Corresponding author. Email: yejun@oist.jp

**Abstract:**
Charge transport in solids at low temperature reveals a material's mesoscopic properties and structure. Under a magnetic field, Shubnikov-de Haas (SdH) oscillations inform complex quantum transport phenomena that are not limited by the ground state characteristics. Here, in elemental metal Cr with two incommensurately superposed lattices of ions and a spin-density-wave ground state, we reveal that the phases of several low-frequency SdH oscillations in $\sigma_{xx}$ ($\rho_{xx}$) and $\sigma_{yy}$ ($\rho_{yy}$) are opposite, contrast with oscillations from normal cyclotron orbits that maintain identical phases. We trace the origin of the low-frequency SdH oscillations to quantum interference effects arising from the incommensurate orbits of Cr's superposed reciprocal lattices, and explain the observed $\pi$-phase shift by the reconnection of anisotropic joint open and closed orbits.

**Text**

Functional materials with interesting and useful electronic, magnetic, and optical responses can be created through the engineering of their electronic structure [1-3]. While conventional crystalline materials still hold many hidden degrees of freedom for unconventional quasiparticles of topological nature [1], an alternative route to access complex electronic behavior is to artificially superpose lattices upon each other to avoid the confinement of the three-dimensional space group symmetry [2-3]. Examples include the control of the termination layer at the interface of two insulating oxides [2] and the twisted overlay of two graphene sheets at small angles [3]. New properties peculiar to the composite lattice can emerge, such as superconductivity [4-5].

These examples suggest the power of exploring degrees of freedom beyond the examples cited above to construct unique types of composite lattices. One natural, but largely unexplored example exists in incommensurately modulated crystalline materials [6]. For conventional crystals, the electronic structures are dictated by their 3D space groups [1]. The symmetry property of an incommensurate structure, on the other hand, is mathematically constructed from space groups of a higher dimensional space, before being sectioned into three dimensions [7]. Many incommensurate structures exist in metals, such as helical magnets and spin and charge density waves [6, 8, 9]. The incommensurate superstructure involving either charge or spin introduces a second set of reciprocal lattices which interacts with the underlying ions. While historically treated as simply opening a gap at the Femi surface of the first set of reciprocal lattices [10-12], multiple applications of the superposed incommensurate reciprocal lattice vectors can provide for more complex possibilities and properties. As we detail below, sophisticated galvanomagnetic behavior emerges from such composite electronic structures because of the incommensurate nature of their superposition.

We explore unconventional electronic characteristics in the archetypal spin-density-wave (SDW) system, Cr. Cr possesses a simple body-centered cubic Bravais lattice and a one-atom basis, which allows a high-fidelity theoretical understanding of its paramagnetic band structure. The paramagnetic Fermi surface is composed of only closed forms with no open sheets and is isomorphic to those of W and Mo (Fig. 1A) [8, 13]. Below $T_N$ = 311.5 K, long-range SDW order with an incommensurate wavevector **Q** = (0.952, 0, 0) develops in Cr as the result of a nesting instability at the Fermi surface (Fig. 1) [8, 12]. Because of different sizes of the hole and electron octahedra, the paramagnetic Fermi surface is imperfectly gapped with residual fragments of both carrier types (Fig. 1).

We find that the superposed reciprocal lattices of the ionic lattice and the SDW lead to two low-frequency SdH oscillations of 36 T and 40 T, with different galvanomagnetic behavior in field dependence, anisotropy, and existence of harmonics. Despite their differences, the SdH oscillations at each frequency reliably demonstrate opposite phases (or equivalently, a π-phase shift) between two configurations of electrical current **I** flowing either parallel or perpendicular to **Q**. In each case, **I** is perpendicular to the external magnetic field **H**. For both the SdH and dHvA oscillations, their tiny areas, in the range of 260 to 440 ppm of the first Brillouin zone cross section, represent a construction through multiple applications of wavevector **Q** over the fragmented original Fermi surface, a mechanism generic for metals with incommensurate superlattices. We attribute these sub-50 T SdH oscillations to a quantum interference effect in a fractal network of orbits. The opposite phase condition between the different resistivity channels is explained by further considerations of anisotropic conduction in joint open and closed orbits in the electronic structure, moving beyond the simple application of either conventional or Berry-phase theories. This novel quantum transport phenomenon should be able to be engineered more generally in appropriately configured quantum and topological materials.

**Single Q-domain of SDW Cr.** Despite the three-fold degeneracy of **Q** under the cubic symmetry, a proper and verifiable field-cooling procedure reliably induces a single-Q domain along one of the cubic axes, with a volume ratio greater than 98% [14-16], Methods) (Fig. 2). The state of a single Q-domain allows measurement conditions to be specified uniquely by the configuration of **Q**, electrical current **I**, and magnetic field **H** (Fig. 2). For a single crystal specimen wired in a standard six-lead configuration for both MR and Hall measurements (Fig. 2, Methods, Supplementary Fig. 1), switching the single Q-domain between different cubic axes provides a technical advantage to fully access the resistivity matrix $\rho(H_z) = \begin{bmatrix} \rho_{xx} & \rho_{xy} \\ \rho_{yx} & \rho_{yy} \end{bmatrix}$ in a coordinate system defined by $\mathbf{Q} \parallel x$, and $\mathbf{H} \parallel z$, as the fixed lead placement largely avoids systematic errors. We follow this convention of x-y-z from here on. Hall resistivities $\rho_{xy}$ and $-\rho_{yx}$, measured under $\mathbf{Q} \parallel \mathbf{I}$ and $\mathbf{Q} \perp \mathbf{I}$ configurations, respectively, provide the experimental verification of our single-Q procedure by confirming the current-voltage reciprocity $\rho_{xy} = -\rho_{yx}$ (Fig. 2). Our measured $\rho_{xy}$ and $-\rho_{yx}$ are consistent with results reported in the literature under the identical **H-I-Q** configuration [14, 17], and the non-monotonic field dependence indicates carriers of different mobilities transitioning from limits of low to high field.

As the **Q** wavevector breaks the four-fold symmetry in the *x-y* plane, a large anisotropy exists between $\rho_{xx}$ and $\rho_{yy}$ even at zero field [16]. The contrast in our measured resistivities $\rho_{xx}$ and $\rho_{yy}$ (Fig. 2B) is more pronounced than that reported in the literature [16], testifying to the completeness of our single Q-domain state (Methods). Under field at *T* = 1.65 K, the anisotropy remains between $\rho_{xx}(H_z)$ and $\rho_{yy}(H_z)$ (Fig. 2C), which is consistent with what is reported in the literature between $\mathbf{Q} \parallel \mathbf{I}$ and $\mathbf{Q} \perp \mathbf{I}$ [14].

**π-phase shifts between SdH oscillations.** From all four matrix elements of $\rho(H_z)$ at 1.65 K, SdH oscillations are extracted and plotted vs. 1/*H* in Fig. 3. For many SdH frequencies, the phase relationship in $\rho_{xx}$, $\rho_{yy}$, and $\rho_{xy}(-\rho_{yx})$ can be directly visualized and compared. There are two types of behavior. For high frequency (1258 T), the phases of SdH oscillations in $\rho_{xx}$ and $\rho_{yy}$ are identical (Fig. 3B). However, for low-frequency (36 T) SdH oscillations, there exists an unusual opposite phase relationship (or a π-phase shift) between $\rho_{xx}$ and $\rho_{yy}$, which can be seen directly by inspecting the raw data in Fig. 3A. Through fast Fourier transform (FFT) analysis (Fig. 4), we discover that there exist both 36T and a set of 40/80/120T frequencies in each element of $\rho$, with 80 T and 120 T being the harmonics of 40 T. While the opposite phase relationship of the 120 T oscillation is clearly observable in the raw data (Fig. 3C), it is difficult to directly visualize the 40 and 80 T oscillations in $\rho_{yy}$. Here we extract precise phases of several of the strongest SdH oscillations through direct fitting of the raw SdH data (Methods, Supplementary Fig. 3). Overall, all four frequencies of 36/40/80/120T have a π-phase shift between $\rho_{xx}$ and $\rho_{yy}$, while the phases at 1258T are identical (Supplementary Fig. 3). For $\rho_{xy}(-\rho_{yx})$, it is visually verifiable that the phase of the 36T oscillations matches that of $\rho_{yy}$, while the phase of the 120 T oscillations is identical to that of $\rho_{xx}$ (Fig. 3D).

Our discussion of SdH oscillations has to this point focused on the resistivity matrix $\rho$, instead of the conductivity matrix $\sigma$ [18]. A full calculation of $\sigma$ verifies that the π-phase shifts between SdH oscillations in $\rho_{xx}$ and $\rho_{yy}$ also exist between $\sigma_{xx}$ and $\sigma_{yy}$ (Supplementary Fig. 2). Indeed, we have a relatively simple situation in three-dimensional Cr; both $\rho_{yy}$ and $\rho_{xx}$ in Cr dominate over $\rho_{xy}$ (Fig. 2). As $|\rho_{xy}\rho_{yx}/\rho_{xx}\rho_{yy}| < 6\%$ at all fields, we have $\sigma_{xx} \sim 1/\rho_{xx}$ and $\sigma_{yy} \sim 1/\rho_{yy}$ [18]. Moreover, given the conventional phase relationship of the 1258 T SdH oscillations between $\rho_{xx}$ and $\rho_{yy}$ (Fig. 3A), the opposite phases of the 36/40/80/120 T oscillations between *x* and *y* channels cannot be attributed to the matrix inversion. Instead, the data reveal the genuine characteristics of a novel quantum transport phenomenon.

We have made additional checks by measuring the magnetoresistivity $\rho_{(110)}$ from a different piece of a Cr single crystal of a different origin, with **I** ∥ (1,1,0), **Q** ∥ (1,0,0), and **H** ∥ (0,0,1) (Fig. 3, Methods, Supplementary Fig. 1B). Both the 36 T and 40 T SdH oscillations are confirmed in $\rho_{(110)}$, together with the higher harmonics at 80 T and 120 T, and other ordinary frequencies (Figs. 3-4, Supplementary Table 1). From the raw SdH oscillations (Fig. 3), the 36.0 T frequency can be verified to a precision within ±0.2 T. The observed SdH oscillations reveal that the phases of the 36, 40, 80, and 120T orbits remain identical to those of $\rho_{xx}$ (Fig. 3). Hence, the phases of all these low-frequency oscillations should follow one of the binary values in $\rho_{xx}$ and $\rho_{yy}$, rather than continuously varying over $\pi$. The MR measured along all other directions in the *x-y* plane can be considered as a linear combination of $\rho_{xx}$ and $\rho_{yy}$.

We show the temperature evolution of $\Delta\rho_{xx}(H_z)/\rho_{xx}$ in Fig. 4. The four low-frequency oscillations dominate the SdH response, as at $T$ = 1.65 K. The $\Delta\rho/\rho$ amplitudes of all high-frequency SdH oscillations (> 150 T) in $\rho_{xx}$ are weaker than that of the 36T mode by at least a factor of 10 (Fig. 4B). All other SdH oscillations disappear below our sensitivity level at ~12 K (Fig. 4B), while the 40 T family of oscillations disappears at ~70 K in $\rho_{xx}$ (Figs. 4A, 4C), and the 36 T oscillation remains noticeable at 90 K (Fig. 4A).

Despite being close in frequency, the 36 T and 40 T SdH oscillations have very different characteristics, which allows an experimental separation of the two. At 1.65K, the 36 T mode has comparable amplitude ratios between $\Delta\rho_{xx}/\rho_{xx}$ (1.1 10$^{-3}$ in Fig. 4b) and $\Delta\rho_{yy}/\rho_{yy}$ (8 10$^{-4}$ in Fig. 4D). It is noticeable from 1.5 Tesla up with a mild field dependence and no detectable higher harmonics (Figs. 3-4). By comparison, the 40 T oscillation is always accompanied by strong harmonics at 80 T and 120 T and all three are only observable above ~7 T in our samples (Figs. 3-4). The 40 T oscillation and its harmonics have strong amplitudes in both $\rho_{xx}$ and $\rho_{xy}(-\rho_{yx})$ but are weak in $\rho_{yy}$.

**dHvA oscillations.** The paramagnetic Fermi surface of Cr is isomorphic to those of Mo and W, as all are composed of closed forms and differ in energy only by the varying strength of spin-orbit coupling [19-21]. Under the same configuration of **H** ∥ (0,0,1), both Mo and W have multiple dHvA frequencies in the range of 10000 to 25000 T from the large hole and electron octahedra, but no dHvA frequency below 500 T [19-21]. For Cr in the single-Q SDW state under the **Q** ⊥ **H** configuration, there is no dHvA quantum oscillation frequency higher than 4000 T (Fig. 5A). Instead, Refs. [15, 22] reported that two frequencies under 50 T emerge (Supplementary Table 1), indicating that the large octahedral Fermi surfaces are partially gapped in the SDW state and the sub-50 T frequencies are created by the **Q** vector.

Our precise control of the single-Q domain state provides the opportunity to measure dHvA oscillations in comparison with their SdH counterparts (Methods). We have detected dHvA oscillations on two Cr samples (Fig. 5A, Methods, Supplementary Fig. 1), and the frequencies above 150 T are consistent with reported values in the literature [15, 22] (Supplementary Table 1). Most importantly, we reveal several low-frequency dHvA oscillations at 26, 32, and 44 T (Fig. 5, Methods). Our measured SdH and dHvA frequencies are consistent within each technique, over multiple samples and multiple cooldowns of each. The differences between sub-50T quantum oscillations observed by dHvA and SdH techniques are larger than the experimental uncertainty, and represent a real discrepancy.

A study of the temperature dependence of the dHvA oscillations reveals further differences. At 12K, amplitudes of all dHvA frequencies diminish below 5% of that of 1.65K, and all sub-100T modes are no longer detectable (Fig. 5B). At 25 K, only the 428 T and 899 T dHvA oscillations remain noticeable at a level of 0.37% and 0.06%, respectively to their amplitudes at 1.65K. This behavior sharply contrasts with that of the SdH modes as they disappear above 90K and 60K, respectively (Fig. 3). Our measurements suggest that the SdH oscillations of 36T and 40T (Fig. 3) are unlikely to share the same orbits of the dHvA oscillations (Fig. 5).

**Q-induced composite dHvA/SdH orbits.** No low-frequency SdH/dHvA oscillations are expected to arise from the paramagnetic Fermi surface. Hence we first discuss generic schemes of how such orbits can be constructed through the wave vector **Q**, using the residual fragments of the Fermi surface that remain after electron and hole octahedra have been imperfectly gapped by the SDW [15, 22, 23]. Historically it has been assumed that either the electron octahedral surface was fully destroyed by the incommensurate SDW or, alternatively, a **Q**-induced band structure reconstruction was applied to hole ellipsoids that do not relate to the SDW gap, leading to much larger suggested SdH/dHvA frequencies of 100-200 T. Our measured SdH/dHvA periods between 26-44 T represent a cyclotron cross section $S$ of 2.5-4.2 $10^{-3}$ Å$^{-2}$, which is about 260-440 ppm of the first Brillouin zone cross-section area ($8\pi^2/a^2$ with the *bcc* lattice constant *a*=2.882Å). Connected by only the primary **Q** vector, a potential orbit can be formed through residual fragments of the nested octahedra along the cubic axis (1, 0, 0) (Fig. 1D), which has been referred to as the "banana" orbit [10-11]. As the lateral sizes of dHvA/SdH orbits are about 0.02 r.l.u., while $(1 - Q) \cong 0.05$ r.l.u., other types of orbits can be constructed through a combination of shiftings by multiples of **Q**. For example, an orbit shaped as a "triangle" is formed by an electron segment in addition to two hole-segments displaced relatively by 2**Q**, opening a gap at point $v$ to avoid band crossing (Fig. 1E). On these orbits, charge carriers travel alternately from electron to hole segments, circling a closed loop connected by the primary path and multiples of **Q**, thereby creating low-frequency dHvA oscillations.

Schemes of Fermi surface reconstruction are complicated by the unknown sizes of Fermi surface fragments after the SDW gap is created. Unfortunately, an existing photoemission study on Cr films [24] did not clearly resolve the gapped area and did not have a sufficiently fine resolution in reciprocal space to define the Fermi surface fragment to the degree needed here. What we do know is that the dHvA oscillations of 26-44T have surprisingly large amplitudes, about 30% of the strongest amplitude of 899 T (Fig. 5A). These large dHvA amplitudes of closely spaced frequencies imply a large Landau level DOS from many degenerate orbits in the re-constructed electronic structure.

We further note that the sub-50T SdH oscillations are different in frequency from the dHvA oscillations and survive to the higher temperatures of 60-90 K as compared to the 12 K of the dHvA oscillations. Taken together, we deduce that instead of a conventional interpretation of extremely light carriers, these SdH features are likely created by quantum interference effects [25-27]. The quantum interference introduces SdH oscillations that do not have dHvA correspondences and labels carriers with seemingly low masses.

Because the Q-vector is incommensurate with the paramagnetic unit cell, the existence of 'orbits' as shown in Fig 1 is a simplification. In the absence of scattering, a quasiparticle on the Fermi surface will peregrinate in an aperiodic fashion and orbits will never close. Effective closed orbits (such as Fig. 1D) and periodic orbits (such as Fig. 1E) result from an approximation where only the lowest-order coupling of a quasiparticle to the SDW is taken into account. Because SdH is a resistive (scattering) measurement, the effective orbits observed will be weighted by the mean free time before momentum non-conserving (impurity, phonon) scattering resets the quasiparticle momentum into the rest frame. In contrast, dHvA is a thermodynamic property and is weighted by the density of states; impurity scattering will broaden what is in principle a fractal-like forest of peaks and the dominant frequencies will not be identical to those seen in SdH. It is the reconnection of orbits via scattering or tunneling processes (loosely labelled "magnetic breakdown" [28]) that allows periodic behavior to emerge. As we now show, the important influence of reconnections is manifested in the strong anisotropy, and the pi phase shift.

**An anisotropic origin of the $\pi$-phase shift.** Both dHvA and quantum interference oscillations represent quantization by a nonlocal quantity of loop size [25-26]. Therefore, SdH oscillations from two orthogonal channels, $\rho_{xx}(\sigma_{xx})$ and $\rho_{yy}(\sigma_{yy})$, are not expected to differ by a random phase between 0 and $\pi$. Instead they likely would have opposite phases or equivalently a factor of $e^{i\pi} = -1$, a $\pi$-phase shift. This discrete distribution of SdH phase is supported by the experimental observation in $\rho_{(110)}$. The opposite phase scenario is further supported by the harmonic series 40/80/120T given that all have a $\pi$-phase shift, instead of phase shifts of $\pi/2\pi/3\pi$ (Supplementary Fig. 3).

Here we explain this $\pi$-phase shift from a semi-classical perspective. We first consider the universal features of open orbits in the superposed lattices. The repetitive application of **Q** in Fig. 1E induces both the closed "triangle" orbit and a generic open orbit "…-v-v-v-…". Similar open orbits were constructed from hole ellipsoids and posited to account for the resistivity anisotropy between $\rho_{xx}$ and $\rho_{yy}$ [14]. Here, with the open orbit parallel to **Q** along the $x$-axis (Fig. 2A) and the carriers' velocity orthogonal to the orbital direction, conduction along the open orbit is independent of $H$ and dominates $\sigma_{yy}$ as $\sigma_{yy} \sim \sigma_{yy,oo} = const$. Other closed orbits only make minor contributions to $\sigma_{yy}$ that diminish at high field as $\sigma_{yy} = \sigma_{yy,oo} + \sigma_{yy,co} = const. + AH^{-2}$. Therefore, $\rho_{yy}(H)$ only depends weakly on $H$ and eventually saturates (Fig. 2C). By contrast, the open orbit has no effect on $\sigma_{xx}$ ($\rho_{xx}$). Magnetoresistivity from closed orbits alone generically evolves with a strong field dependence as $\rho_{xx} \sim H^2$ in the Lifshitz-Azbel-Kaganov framework [14, 29-30]; the exponent two is expected in the high field limit. Here our $\rho_{xx} \sim H^{-1.3}$, mainly because sharp corners of DW systems produce a linear form of magnetoresistance [31]; our Cr crystals in fields of 14 T are still in the low-field region.

A full theoretical treatment of coupled (coherent) quantum orbits and (incoherent) semiclassical galvanomagnetic processes remains challenging because quantum phases along various branches are dependent on the gauge choice of the vector potential **A** [29]. Here we construct a composite structure in Fig. 6A, featuring an open orbit, a closed orbit, both semiclassical, and their connection region which incorporates a coherent network of small orbits or segments. The small orbits or segments can arise from remnants of Fermi surfaces gapped and translated by the different harmonics of the SDW potential as well as the lattice potential and are difficult to pin down accurately as mentioned above. Such a quantum mechanical connection region can be made equivalent to an effective magnetic breakdown junction [28], with its tunneling probability $P$ (see below) oscillates with $H^{-1}$ due to either Landau quantization or quantum interference [25] in the small coherent network [29]. The salient point is that, for $\rho_{xx}$ the galvanomagnetic response is dominated by the closed loop, while for $\rho_{yy}$ the galvanomagnetic response is dominated by the open orbit. The carriers of the closed loop are the leading component of conduction in $\rho_{xx}$, but only have a perturbative influence on $\rho_{yy}$ through the connection region to the (nearly) free carriers along the open orbit. This difference in the identity of the dominant conduction carriers can explain opposite phases in the SdH oscillations.

Carriers along the four branches in Fig. 6A, flowing in and out of the quantum region, satisfy the microbalance condition. Because of the closed orbit, currents on branches 3 and 4 are identical under steady-state conditions, and we can denote a probability $P$ between 0 to 1 for carriers tunneling between branches 1 and 4, and simultaneously between branches 3 and 2. Conductivity changes according to $P$ can be understood through the perspective of an effective $\omega\tau$, which governs the galvanomagnetic responses. We define cyclotron angular frequencies $\omega$ for both the closed (*co*) and open orbits (*oo*) with $\omega_{oo} \ll \omega_{co}$. For a change in the coupling $\delta P > 0$, more carriers interact between the two orbits and the open orbit has an effective increase of $\omega_{oo}$. The overall conduction $\delta\sigma_{yy} < 0$. At the same time, more carriers in the closed orbit experience segments of the open orbit, causing a reduced $\omega_{co}$, and correspondingly $\delta\sigma_{xx} > 0$. As $P$ oscillates because of the quantum interference effect within the connection region, $\sigma_{xx}$ and $\sigma_{yy}$ have opposite phases in their SdH signatures. Numerically simulated SdH oscillations are presented in Fig. 6B for all independent galvanomagnetic channels. More details of the theoretical framework for solving the coupled orbits problem and of its application to the present model are provided in the Supplementary Note. We note that in this model SdH oscillations in the Hall conductivity have either a 0 or $\pi$ phase shift to their counterparts in the magnetoresistance channels, instead of the $\pi/2$ difference in the integer quantum Hall effect.

The SdH oscillations of 36T and 40/80/120T frequencies are largely different in their galvanomagnetic behavior, for aspects such as harmonics, the field range, and amplitude anisotropy between $\rho_{xx}$ and $\rho_{yy}$. Yet they all demonstrate an opposite phase between two conduction channels. The insensitivity to details can be explained by the generic mechanism described above. Their differences can be attributed to microscopic differences in the semiclassical orbits coupled through different quantum regions (Supplementary Note). For example, for SdH oscillations in the Hall conductivity, the fact of carriers being mainly of electron or hole type can shift the phase by 0 or $\pi$.

Conventional SdH oscillations reflect the Landau levels' density of states [26] and those of a linearly dispersive (Dirac) band structure would incorporate an extra $\pi$-phase due to the Berry phase [32]. In both scenarios, with the current flowing perpendicular to the field, $\sigma_{xx}$ and $\sigma_{yy}$ are expected to have identical phases in the SdH oscillations. Importantly, our observed low-frequency SdH oscillations in Cr do not follow these simple scenarios. Instead, the opposite-phases between $\rho_{xx}(\sigma_{xx})$ and $\rho_{yy}(\sigma_{yy})$, contrasted by their congruence in the conventional 1258 T SdH orbit, reflect an intrinsically new, anisotropic mechanism of quantum transport. The two incommensurately superposed lattices in Cr naturally create the quantum interference effect, which, in principle, could be crafted in related quantum and topological materials.

**Acknowledgments**
We thank M. G. Dronova and D. M. Silevitch for technical assistance during dHvA measurements. **Funding:** Y. F. acknowledges the support from Okinawa Institute of Science and Technology Graduate University with subsidy funding from the Cabinet Office, Government of Japan. Y. W. acknowledges the startup support from the University of Tennessee, Knoxville. T.F.R. acknowledges support from the U.S. Department of Energy Basic Energy Sciences Award No. DE-SC0014866 for the data analysis and manuscript preparation performed at Caltech. H.C. acknowledges support from U.S. NSF CAREER grant DMR-1945023. **Author contributions:** Y. F. designed research. All authors performed research and analyzed data. Y. F. prepared the manuscript and all authors commented. **Competing interests**: The authors declare no competing interests.

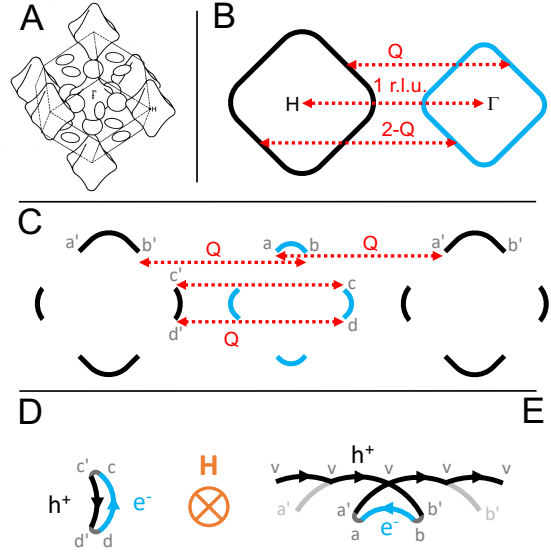

**Fig. 1. Fermi surface and cyclotron orbits of two superposed reciprocal lattices.**
(**A**) Calculated paramagnetic band structure of Cr, which is composed of closed octahedral and ellipsoidal forms of hole and electron pockets, is similar to those of Mo and W. The schematic is adapted from Ref. [13]. (**B**) A schematic of octahedral Fermi surfaces of hole (black) and electron (blue) projected onto the *a-b* plane. Their parallel surfaces are matched through the nesting wave vector **Q**. (**C**) Upon the formation of a SDW state, the long-range antiferromagnetic order introduces a second set of reciprocal lattices, with the wave vector **Q** serving as the basis. **Q** connects fragmented electron and hole Fermi surfaces in the original ionic reciprocal lattice into continuous forms in panels (D) and (E). (**D**) Under field, carriers flow along the "banana" orbit "*c-c'-d'-d-c*", as tips of the electron and hole octahedra are not annihilated by the nesting condition. (**E**) Another "triangle" orbit is formed when two pieces of large hole arc (*a'-b'*) and a small electron arc (*a-b*) are connected into a closed loop "*b-a-a'-v-b'-b*". Here two hole-arcs "*a'-b'*" are relatively displaced by 2**Q**, and a gap opens at the band crossing point *v*. A full superposition of the two reciprocal lattices would create an open orbit "…*v-v-v-v-v*…".

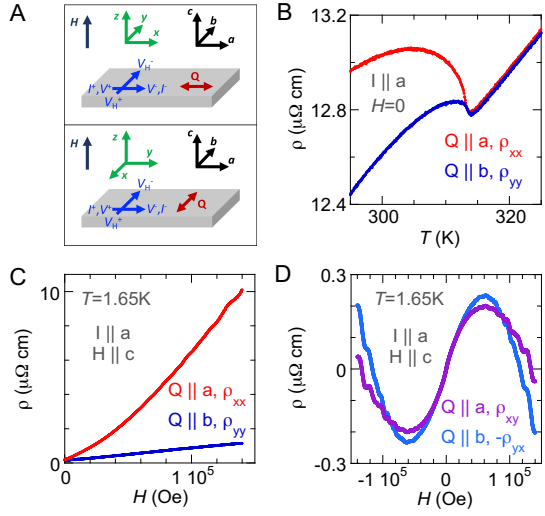

**Fig. 2. Full resistivity matrix $\rho$ measured through single-Q-domain control.** (**A**) A bar-shaped single crystal Cr sample (grey), with the cubic lattice structure marked by the coordinates "*a-b-c*" (black), is wired under a six-lead configuration for both Hall and MR measurements (blue marking); the real samples are listed in Supplementary Fig. 1. Separate single Q-domains (red) along two different cubic axes (Methods) permits us to measure the full resistivity matrix $\rho$ in the **H ∥ z** geometry with a fixed lead placement. Coordinates *x-y-z* (green) of $\rho$ are always defined by orientations of **Q** and **H** as **Q ∥ $x$** and **H ∥ z**. (**B**) Zero-field resistivities of Cr in a single-Q state are contrasted near $T_N$ between **Q ∥ I** (red) and **Q ⊥ I** (blue) configurations and can be compared to the literature [16]. (**C**) Magnetoresistivities $\rho_{xx}(H_z)$ and $\rho_{yy}(H_z)$ reveal a large anisotropy, yet both are much larger than (**D**) the Hall resistivities $\rho_{xy}(H_z)$ and $-\rho_{yx}$. The consistency between Hall resistivities $\rho_{xy}$ and $-\rho_{yx}$ verifies our control of a single-Q state as the measurements were performed with **Q** along two different axes with a constant lead configuration. The specific **I-Q-H** configurations for measured $\rho$ matrix elements are individually listed in the panels.

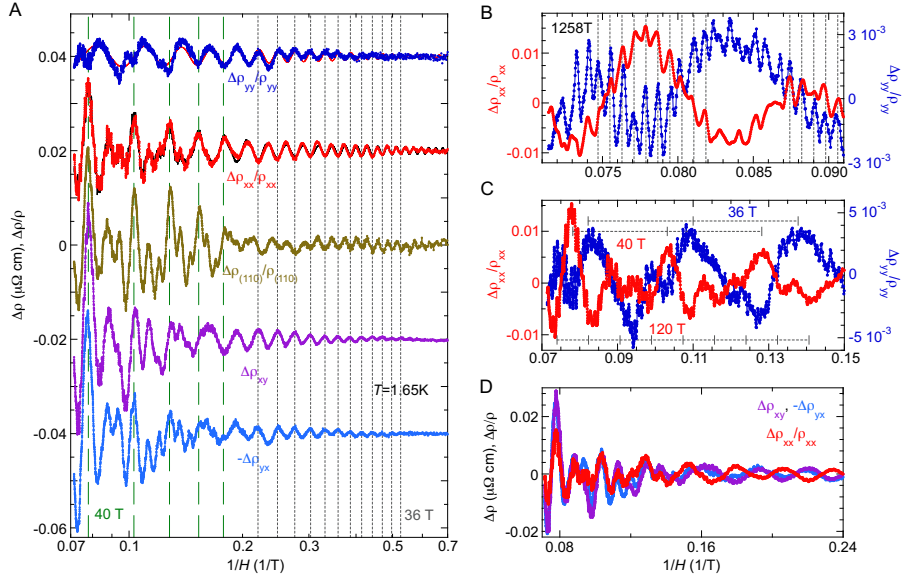

**Fig. 3. π-phase shifts between SdH oscillations of $\rho_{xx}$ and $\rho_{yy}$.** (**A**) Raw SdH oscillations of $\Delta\rho_{xx}/\rho_{xx}$, $\Delta\rho_{yy}/\rho_{yy}$, $\Delta\rho_{(110)}/\rho_{(110)}$, $\Delta\rho_{xy}$, and $-\Delta\rho_{yx}$, vertically displaced for clarity, are plotted versus $1/H$. A log-*x* scale is used to demonstrate the high field behavior. Two sets of vertical lines mark the periodicities of 36 T (grey short dash) and 40 T (green long dash) oscillations which are prominent in different regions of field. For $H < 6$ T, all SdH oscillations are dominated by the 36 T mode. The 40 T frequency is very prominent with $\Delta\rho_{(110)}/\rho_{(110)}$, together with the 120 T oscillations in between. $\Delta\rho_{yy}/\rho_{yy}$ is fit with a single frequency of 36T with an exponentially-decaying amplitude (red line) to demonstrate its dominant presence in that channel. $\Delta\rho_{xx}/\rho_{xx}$ is fit with six frequencies (Methods) to demonstrate the extraction of phases of the SdH oscillations. The phase of the 36 T oscillation in $\rho_{yy}$ is opposite to those of $\rho_{xx}$ and $\rho_{(110)}$, but identical with those of $\rho_{xy}$ and $-\rho_{yx}$. (**B**) For a normal orbit of 1258T, SdH oscillations in $\rho_{xx}$ and $\rho_{yy}$ are in phase, as indicated by the set of vertical dashed lines that are separated by the oscillation period. (**C**) Periods of 36 T, 40 T, and 120 T oscillations are marked for $\Delta\rho_{xx}/\rho_{xx}$ and $\Delta\rho_{yy}/\rho_{yy}$ in the high field range. With ticks on the scale bar marking the local minima and maxima positions of $\rho_{xx}$ and $\rho_{yy}$, respectively, the 120 T SdH oscillations can be verified visually to be of opposite phase (a π-phase shift). (**D**) Between $\rho_{xx}$ and $\rho_{xy}(-\rho_{yx})$, both the 40 and 120 T SdH oscillations are visibly in phase at high field, but 36 T oscillations between these two channels have a π-phase shift that becomes clear at low field.

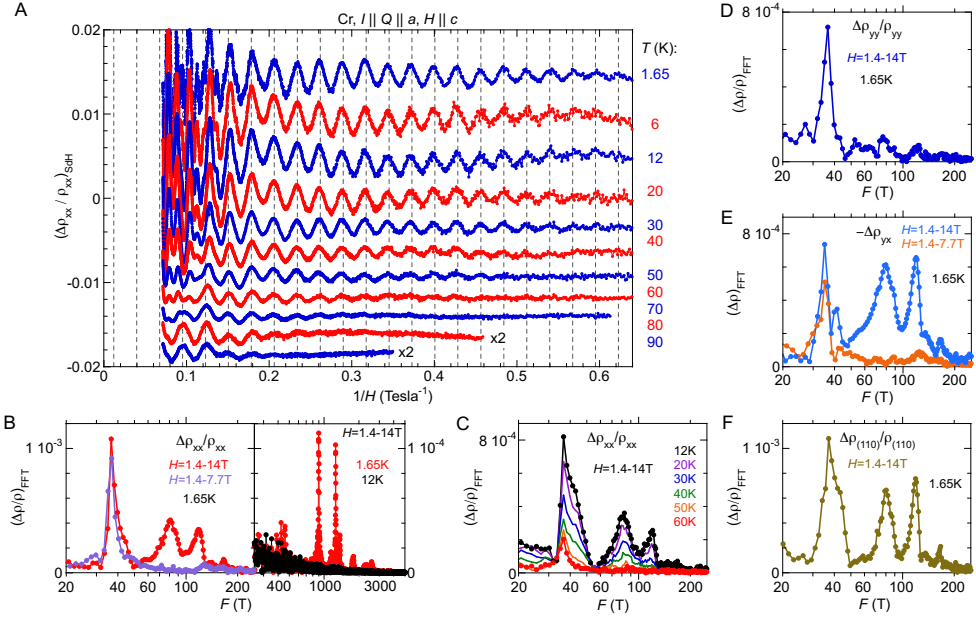

**Fig. 4. Low-frequency SdH oscillations.** (**A**) The temperature evolution of $\Delta\rho_{xx}/\rho_{xx}$ is plotted vs. $1/H$ from $T = 1.65$ to 90 K. All data are separated vertically for clarity. The vertical dashed lines mark the maximal position of $\Delta\rho_{xx}/\rho_{xx}$ in a regular spacing of $1/36.0$ T$^{-1}$. (**B**) FFT spectrum of $\Delta\rho_{xx}/\rho_{xx}$ reveals two SdH frequencies of 36 T and 40 T, and higher harmonics of the latter at 80 T and 120 T. These low-frequency SdH oscillations (left-hand y-scale) dominate higher-frequency quantum oscillations in $\Delta\rho_{xx}/\rho_{xx}$ (right-hand y-scale) by a factor ~10. The 40 T oscillation and its two harmonics exist only in the high field range above 7.7 T, while the 36 T oscillation exists down to ~1.5 T. (**C**) The 36 T and 40 T SdH oscillations exist to high temperature of 90 K and 60 K, respectively; their temperature evolution provides a clean separation. The harmonics 80/120 T share the same temperature evolution as the 40 T oscillation. (**D-F**) FFT spectra of $\Delta\rho_{yy}/\rho_{yy}$, $-\Delta\rho_{yx}$, and $\Delta\rho_{(110)}/\rho_{(110)}$ display characteristics of the 36 T and 40 T oscillations. $\rho_{(110)}(H_z)$ is the MR with current **I** flowing parallel to the (1,1,0) axis, and was measured on a separate piece of Cr single crystal (Supplementary Fig. 1B). It demonstrates that the SdH frequencies of 36 T and 40 T, and harmonics 80 T and 120 T, are repeatable and sample independent.

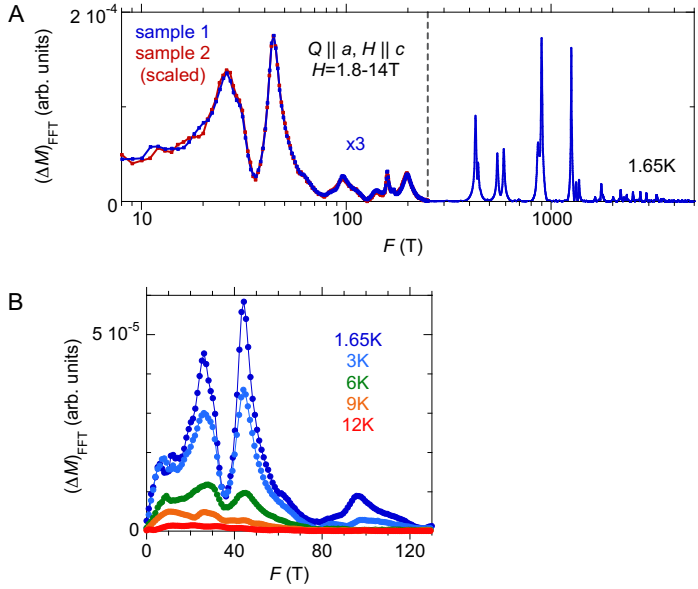

**Fig. 5. dHvA oscillations.** (**A**) The full dHvA spectrum under the specified field-Q configuration. Data from two samples (Methods, Supplementary Figs. 1C, 1D) are consistent, revealing low-frequency modes of 26, 32, and 44 T, in addition to other high frequency modes. Supplementary Table 1 provides a comparison of dHvA frequencies in the literature and our measured dHvA and SdH frequencies. Amplitudes of low-frequency dHvA oscillations are plotted with a scaling factor of 3× for sample 1. The dHvA amplitude of sample 2 is scaled to that of sample 1. (**B**) Temperature dependence of dHvA oscillations of panel (A) at frequencies below 130 T.

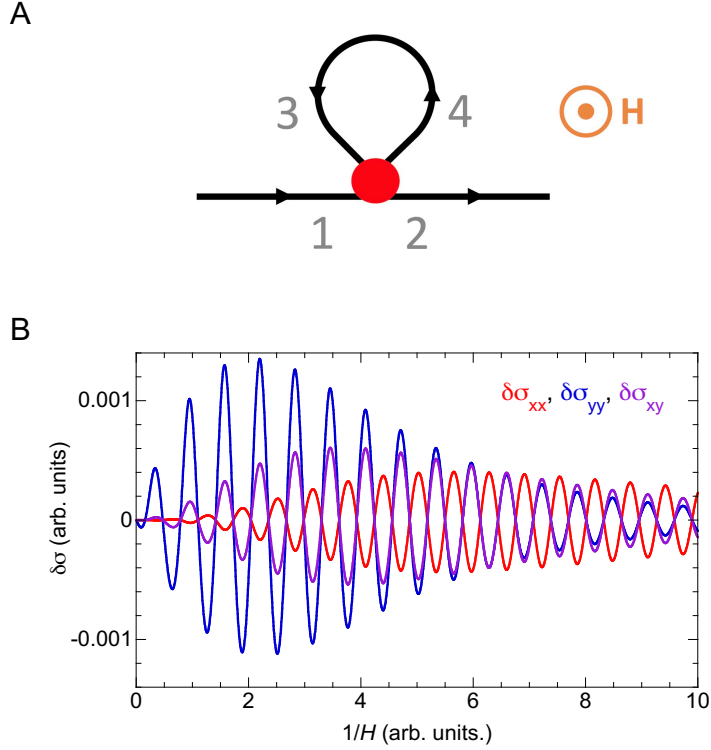

**Fig. 6. Theoretical model for the π-phase shift.** (A) Schematics of a minimal model consisting of semi-classical open (1-2, black straight line) and closed (3-4, black loop) orbits, joined together by a quantum mechanical region (red spot). Both open and closed orbits are responsible for transport under field $H$ of an incoherent (scattering) nature. The connection region is responsible for coherent SdH oscillations arising from quantum interference, which modulates the transition probability $P$ between branches. (**B**) SdH oscillations are simulated using the formulas and parameters given in Supplementary Note for all independent conductivity components of $\sigma$, using the model structure in panel (A). We note $\sigma_{xy} = -\sigma_{yx}$. The opposite phases are clearly demonstrated between $\sigma_{xx}$ and $\sigma_{yy}$. The Hall conductivity $\sigma_{xy}$ has either a 0 or $\pi$ phase difference with respect to $\sigma_{xx}$ and $\sigma_{yy}$, but not $\pi/2$ as in the integer quantum Hall effect.

# Supplementary materials for:

# Quantum interference in superposed lattices

Yejun Feng[1,*], Yishu Wang[2,3], T. F. Rosenbaum[4], P. B. Littlewood[5], Hua Chen[6]

[1]Okinawa Institute of Science and Technology Graduate University, Onna, Okinawa 904-0495, Japan
[2]Department of Materials Science and Engineering, University of Tennessee, Knoxville, Tennessee 37996, USA
[3]Department of Physics and Astronomy, University of Tennessee, Knoxville, Tennessee 37996, USA
[4]Division of Physics, Mathematics, and Astronomy, California Institute of Technology, Pasadena, California 91125, USA
[5]The James Franck Institute and Department of Physics, The University of Chicago, Illinois 60637, USA
[6]Department of Physics, Colorado State University, Fort Collins, Colorado 80523, USA
*Corresponding author. Email: yejun@oist.jp

**Materials and Methods**

**Single Q-state in single crystal Cr.** Specimens of single crystal Cr were procured from two different sources, a wafer of ~2 mm thick and 10 mm diameter with a surface normal of (1, 0, 0) from Alfa Aesar (99.996+%, #13547) [33], and a ~20 mm long boule of ~5 mm diameter from Atomergic Chemetals Inc, Farmingdale, USA. For the Alfa Aesar specimen, the cubic axes within the surface plane were determined by in-lab x-ray diffraction. After the alignment, slices with cubic axes along all surface normals, were cut off from the wafer by a diamond saw. All sawed-off surfaces were polished using 50 nm alumina suspension (MicroPolish, Buehler Ltd.), and the pieces were annealed in vacuum inside sealed quartz tubes at 1000 °C for 32 hrs using a box furnace. After annealing, the pieces were etched to remove surface oxidation using either a Cr etchant (Type 1020, Transene Co. Inc.) [33], or warm 10% HCl solution. The final piece for the electrical transport study had a length of ~5 mm, with a rectangular cross section of $1.84 \times 0.42$ mm$^2$ (Supplementary Fig. 1A). Two smaller plates of Cr were prepared for dHvA measurements, with sizes of $1.82 \times 1.73 \times 0.40$ mm$^3$ and $1.74 \times 1.50 \times 0.20$ mm$^3$, and weight of 10.2 mg and 4.6 mg, respectively. The presence of the SDW/CDW state was confirmed previously by x-ray diffraction studies of other parts of the same wafer [12, 33-34] and the transport signature at $T_N$ (Fig. 2B) [8]. The residual resistivity ratio (RRR) with **I** ∥ **Q**, $\rho(T = 350K)/\rho(T = 2K) \sim 66$. The rod-shaped Cr single crystal from Atomergic Chemetals was indexed by a lab Laue diffractometer. Several slices with a rectangular cross section were cut off from the rod by wire-saw, with the (1, 1, 0) direction along the long direction for the current path, while the side normals are (0, 0, 1) and (1 -1, 0), respectively (Supplementary Fig. 1B). One piece was similarly polished and annealed in vacuum in a quartz tube at 1000 °C for 48 hrs. Surface oxidation was removed by a warm bath of 10% HCl. The final piece measured ~7.5×2.6×0.82 mm$^3$ (Supplementary Fig. 1B).

Both SdH and dHvA measurements were carried out using a horizontal rotator probe inside a 14 T Physical Property Measurement System (PPMS DynaCool-14, Quantum Design, Inc.). The single-Q state is induced by a slow field-cooling (0.4-0.7 K/min) from 350 K down, using a field of 14 T to align the **Q** vector parallel to **H** [14-15]. While there is a small difference in the procedure between Refs. [14-15] as whether to reduce the field to zero at ~150 K above the spin flip transition temperature at ~123 K [35], we find both procedures work. The single-Q state is reflected by measured SdH/dHvA frequencies. Results from both field-cooling procedures are consistent (Supplementary Table 1).

**Shubnikov-de Haas quantum oscillations.** Each of the two galvanomagnetic samples is anchored to the surface of an 8-pin DIP connector using GE varnish (Supplementary Fig. 1). The sample and DIP connector are mounted onto a PPMS rotator sample carrier, which has been modified with differently positioned DIP sockets to accommodate several in-plane angles of field cooling (such as 0°, 90° and 45° for **Q** ∥ **I**, **Q** ⊥ **I**, and **I** ∥ (1,1,0) conditions). Electrical leads (Au wire of 25μm diameter) were attached to the transport samples by silver epoxy (EPO-TEK H20E-PFC, Epoxy Technology) in a six-lead Hall/MR configuration (Supplementary Fig. 1). Both magnetoresistance and Hall resistance were measured using an AC resistance bridge and a preamplifier (LS372 and 3708, Lake Shore Cryotronics, Inc.) for the sample inside the PPMS from $T = 1.65$ to 350 K. An electric current of 10 mA was typically used in order to collect low-noise SdH signals. Each matrix element of $\rho$ was measured under fixed temperature as a function of field. SdH oscillations were extracted as $(\Delta\rho_{xx}/\rho_{xx})_{\text{SdH}} = [\rho_{xx}(H) - \rho_{xx}^{BG}(H)]/\rho_{xx}(H)$ for MR and $\Delta\rho_{xy} = \rho_{xy}(H) - \rho_{xy}^{BG}(H)$ for Hall resistivity, where $\rho^{BG}(H)$ are smooth functions. Measured SdH oscillations from all matrix elements of $\rho$ are plotted vs. $1/H$ in Figs. 3-4, together with the Fourier transformed spectra.

**de Haas-van Alphen quantum oscillations.** Magnetization-based dHvA quantum oscillation measurements were carried out using the torque magnetometry technique (Supplementary Figs. 1C, 1D) based on the Torque Magnetometer option of the PPMS (Tq-Mag, PPMS DynaCool-14, Quantum Design, Inc.). The torque from magnetic anisotropy was detected by piezoresistive elements on the cantilever of Tq-Mag, arranged in a Wheatstone bridge configuration. The change of piezo-resistivity is measured by our own AC resistance bridge (LS372 and 3708, Lake Shore Cryotronics, Inc.). In a torque-based dHvA measurement, the continuous cooling to 1.65 K under the 14 T field can be monitored directly and the preservation of the single-Q state is verified across the spin-flip transition; for electrical transport measurements, there is no resistivity signature of the spin-flip transition [8]. The measured dHvA oscillations, expressed in the unbalanced resistance of the Wheatstone bridge, were extracted after background subtraction and Fourier transformed to the frequency domain. Two pieces of single-Q Cr crystals (10.2 mg and 4.6 mg) were measured to verify the sample and data consistency.

**Fast Fourier Transform.** FFT analysis was applied to both SdH and dHvA oscillations. For dHvA, as the oscillation amplitudes have a strong field dependence, we used a Tukey window function to clean up the FFT spectrum, which is defined as:

$$w(x) = \begin{cases} \frac{1}{2}\left\{1 + \cos\left[2\pi\left(\frac{x}{r} - \frac{1}{2}\right)\right]\right\}, & 0 \leq x < \frac{r}{2} \\ 1, & \frac{r}{2} \leq x \leq 1 - \frac{r}{2} \\ \frac{1}{2}\left\{1 + \cos\left[2\pi\left(\frac{x-1}{r} + \frac{1}{2}\right)\right]\right\}, & 1 - \frac{r}{2} < x \leq 1 \end{cases}$$

In our analysis of dHvA spectra, we chose $r = 0.1$. The Tukey functional form thus affects 5% of the data on the high-field end, as oscillations at the low field end already decay to the noise floor. This minimally affected range of the window function preserves the relative intensities of different oscillation frequencies, while removing interstitial spectral weight and potential resonance types of spectral anomalies. For SdH oscillations, as the signals have only weak field dependence, a straightforward FFT can be applied.

**Phase extraction of SdH oscillations.** For $\rho_{xx}$, $\rho_{yy}$, and $\rho_{(110)}$ at 1.65 K, the raw data of $(\Delta\rho/\rho)_{\text{SdH}}$ was fit to a functional form of $(\Delta\rho/\rho)_{\text{SdH}} = \sum_F A_F \cos(2\pi F/H + \psi_F)\exp(-K_F/H)$, where the oscillation frequencies $F$ are fixed from FFT-analyzed spectra. The summation runs through the six strongest oscillation frequencies $F$, which include 36, 40, 80, 120, 1258 T and either 467 T (for $\rho_{yy}$) or 899 T (for $\rho_{xx}$ and $\rho_{(110)}$). $K_F$ is the decay parameter to capture the exponential field-dependence, similar to the effect of the Dingle temperature.

**Extra References:**

**Supplementary Table 1. dHvA and SdH frequencies of single-Q Cr under the configuration of H || (0,0,1) and Q || (1,0,0).** The dHvA frequencies reported in the literature were extracted from figures in Refs. [Graebner1968, Fawcett1976]. For dHvA, only frequencies with intensities higher than 1% of the strongest peak at 899 T are reported here.

| Extracted dHvA frequencies from the literature (Tesla) | | Measured dHvA frequencies of this work (Tesla) | Measured SdH frequencies from $\rho_{xx}$ (Tesla) | Measured SdH frequencies from $\rho_{(110)}$ (Tesla) |
|---|---|---|---|---|
| Ref. [Graebner1968] | Ref. [Fawcett1976] | | | |
| 32 | 28 | 26 | | |
| | | 32 | 36 | 36 |
| 44 | 41 | 44 | 40 | 40 |
| | | 95 | | |
| | 158 | 158 | 165 | 165 |
| | | 198 | | |
| | 427 | 428 | 426 | 426 |
| | | 440 | | |
| | 564 | 547 | | |
| | | 585 | | |
| | 856 | 862 | | |
| | 903 | 899 | 899 | 898 |
| | 1229 | | | |
| | 1262 | 1256 | 1258 | 1258 |
| | 1340 | 1325 | | |
| | 1376 | 1369 | 1367 | 1368 |
| | 1663 | 1641 | | |
| | 1768 | 1755 | 1752 | |
| | 1808 | 1789 | 1792 | 1793 |
| | 2039 | 2009 | | |
| | | 2186 | | |
| | 2273 | 2268 | 2250 | 2266 |
| | | 2342 | 2352 | |
| | 2392 | | | |
| | | 2511 | 2512 | |
| | | 2727 | | |
| | 2835 | | | |
| | | 2919 | | |
| | | 3278 | | |
| | 3494 | 3447 | | |
| | | 3529 | 3540 | |

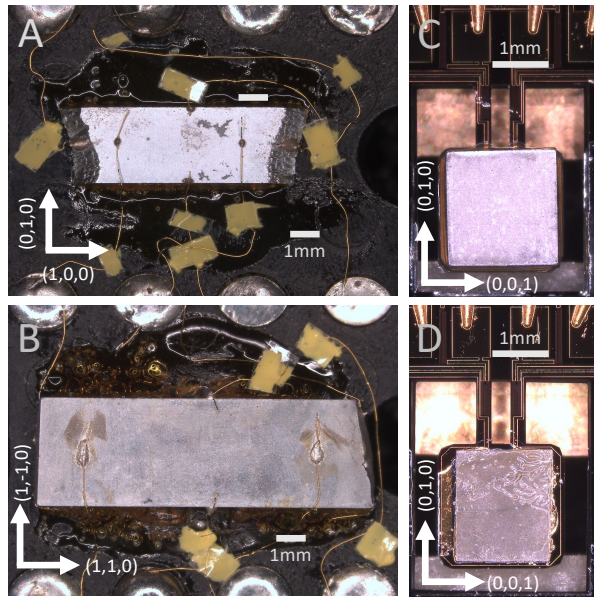

**Supplementary Fig. 1.** Pictures of single crystal Cr samples used in (**A-B**) SdH and (**C-D**) dHvA measurements. The two SdH samples, mounted on top of 8-pin DIP connectors, are from (A) Alfa Aesar, and (B) Atomergic Chemetals, respectively (Methods). The two dHvA samples are from Alfa Aesar, and weigh (C) 10.2 and (D) 4.6 mg, respectively. The crystalline orientations are marked alongside the individual sample.

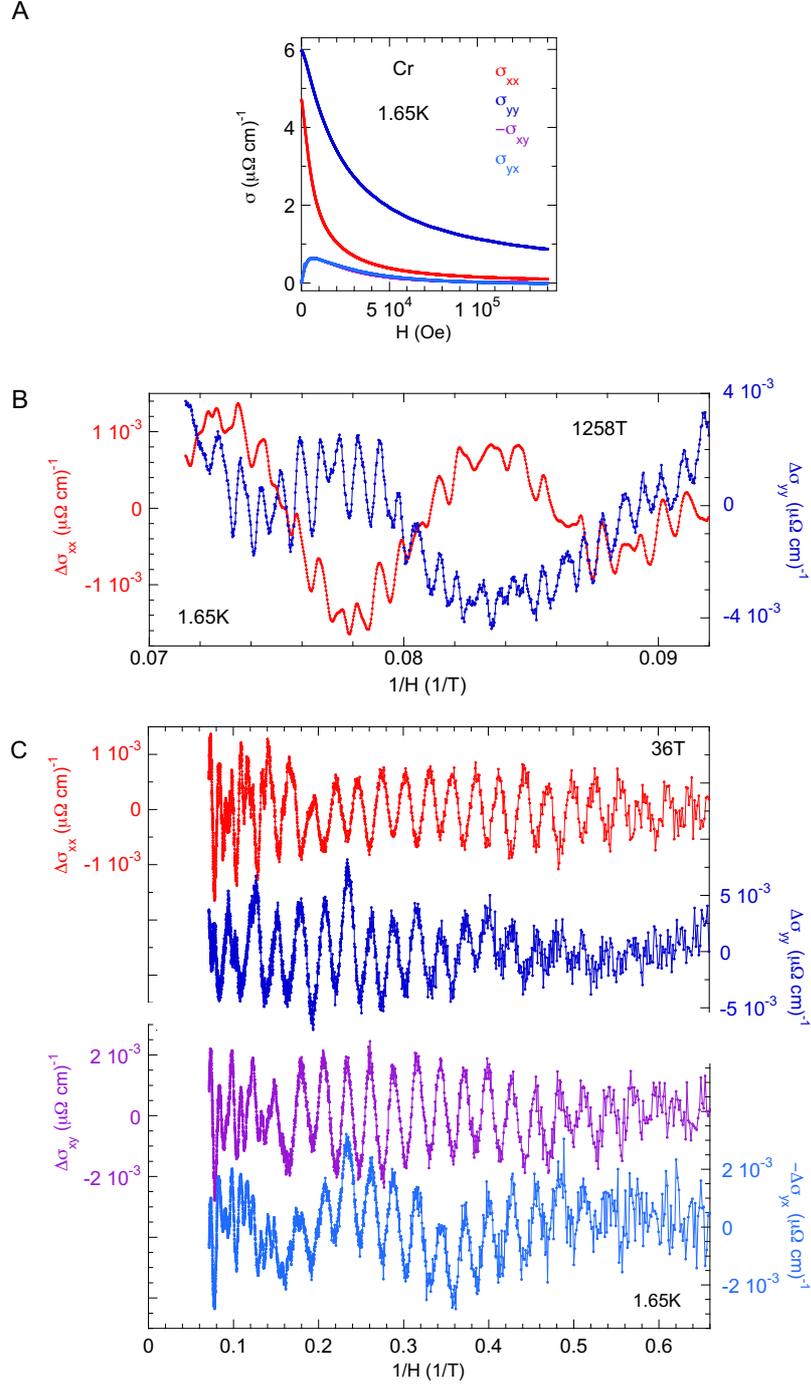

**Supplementary Fig. 2.** (A) Electrical conductivity $\sigma$ calculated from inversion of the resistivity matrix $\rho$ in Fig. 2. (B) 1258 T SdH oscillations in both $\sigma_{xx}$ and $\sigma_{yy}$ have the same phase. (C) 36 T SdH oscillations in $\sigma_{xx}$, $\sigma_{yy}$, $\sigma_{xy}$, and $-\sigma_{yx}$, demonstrating the π-phase difference between $\sigma_{xx}$ and $\sigma_{yy}$. The direct verification of the π-phase difference in $\sigma$ excludes a trivial explanation of matrix inversion from $\sigma$ to $\rho$.

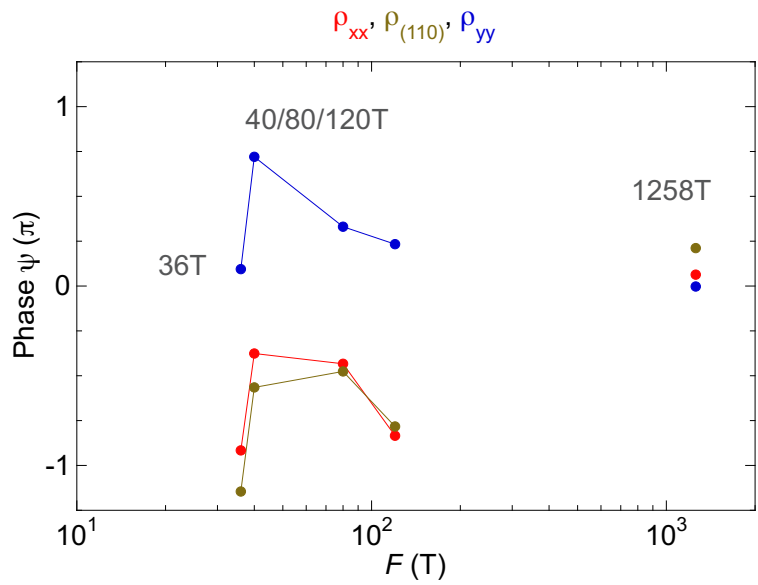

**Supplementary Fig. 3.** Extracted SdH phases $\psi_F$ of 36/40/80/120 T in comparison to those of 1258 T (Methods) are plotted for all three magnetoresistivities $\rho_{xx}$, $\rho_{yy}$, and $\rho_{(110)}$. The π-phase shift is clear for the low-frequency modes.

**Supplementary Note: Semiclassical theory for the π-phase shift**

The semiclassical Boltzmann equation under the relaxation time approximation has the following form in the steady state:

$$\frac{\partial f}{\partial \mathbf{r}} \cdot \dot{\mathbf{r}} + \frac{\partial f}{\partial \mathbf{k}} \cdot \dot{\mathbf{k}} + \frac{f - f_0}{\tau} = 0 \quad (1)$$

where $\tau$ is the relaxation time and $f_0$ is the equilibrium Fermi-Dirac distribution function. Following [30], in the presence of both electric field **E** and magnetic field **B** the Boltzmann equation can be written as

$$e\tau \frac{\partial f_0}{\partial \epsilon} \mathbf{v} \cdot \mathbf{E} - \frac{\partial g}{\partial s} \frac{v_\perp \tau}{\ell^2} = g, \quad (2)$$

where $g \equiv f - f_0$, $\ell = \sqrt{\frac{\hbar}{eB}}$ is the magnetic length, $v_\perp$ is the magnitude of velocity in the plane perpendicular to **B**, and $s$ parameterizes the $k$-space orbit. Eq. 2 has the following solution:

$$g = \int_{-\infty}^{s} ds' \frac{e\ell^2}{v_\perp} \frac{\partial f_0}{\partial \epsilon} \mathbf{v} \cdot \mathbf{E} \exp\left(-\int_{s'}^{s} \frac{\ell^2}{v_\perp \tau} ds''\right), \quad (3)$$

using which one can calculate the electric current and hence the conductivity

$$\mathbf{j} = -\sum_n \int \frac{ds\, d\epsilon\, dk_z}{(2\pi)^3} \frac{e}{\hbar v_\perp} \mathbf{v} g. \quad (4)$$

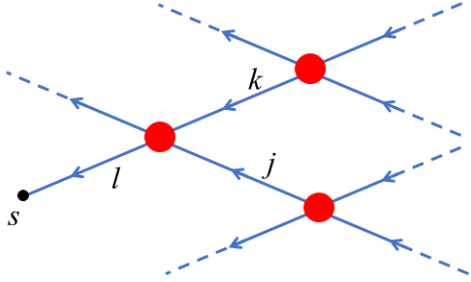

**Figure 1**. Schematics of coupled orbits. Each line represents an orbit segment. Red dots stand for magnetic breakdown junctions. $s$ represents the present position of the electron.

We next discuss how to generalize Eq. 3 to the case of many orbits coupled through magnetic breakdown (MB) junctions [28]. Each MB junction connects two incoming and two outgoing $k$-space paths so that each semiclassical particle tunneling through the MB junction has a nonzero possibility of choosing either of the two outgoing paths. Therefore Eq. 3 must be generalized to a summation over all possible trajectories of a given particle weighted by their respective probabilities:

$$g = \sum_p P_p \int_{-\infty_p}^{s_p} ds'_p \frac{e\ell^2}{v_\perp} \frac{\partial f_0}{\partial \epsilon} \mathbf{v} \cdot \mathbf{E} \exp\left(-\int_{s'_p}^{s_p} \frac{\ell^2}{v_\perp \tau} ds''_p\right) \quad (5)$$

where $p$ stands for path and $P_p$ is the probability of taking a given path. To perform such a path summation, we label all directed segments of the couple orbits by $l, j, k$ etc. as illustrated in Fig. 1. Due to the definite directions of the segments, the starting point of a given segment $l$ can be uniquely denoted by $s_l$, which coincides with the starting point of another segment going out of the same MB junction as $l$. To calculate $g(s)$ with the $s$ indicated in the figure, note that the last part of all paths coincide:

$$g(s) = \int_{s_l}^{s} ds' F_1(s') e^{-\int_{s'}^{s} F_2(s'') ds''} + e^{-\int_{s_l}^{s} F_2(s'') ds''} I_l \quad (6)$$

where

$$F_1 \equiv \frac{e\ell^2}{v_\perp}\frac{\partial f_0}{\partial \epsilon}\mathbf{v}\cdot\mathbf{E}, \qquad (7)$$

$$F_2 \equiv \frac{\ell^2}{v_\perp \tau}$$

$$I_l \equiv \sum_p P_p \int_{-\infty_p}^{s_l} ds'_p F_1(s'_p) e^{-\int_{s'_p}^{s_l} F_2(s''_p)ds''_p}.$$

The key is therefore to calculate $I_l$. However, note that there are only two segments ($k,j$ in Fig. 1) that go into the MB junction from which $l$ leaves. Assuming $I_l$ is well-defined for all $l$, we have

$$I_l = P_{l\leftarrow k}\left[\int_{s_l-s_{Bk}}^{s_l} ds' F_1(s') e^{-\int_{s'}^{s_l} F_2(s'')ds''} + e^{-\int_{s_l-s_{Bk}}^{s_l} F_2(s')ds'} I_k\right] \qquad (8)$$
$$+P_{l\leftarrow j}\left[\int_{s_l-s_{Bj}}^{s_l} ds' F_1(s') e^{-\int_{s'}^{s_l} F_2(s'')ds''} + e^{-\int_{s_l-s_{Bj}}^{s_l} F_2(s')ds'} I_j\right]$$

where $P_{l\leftarrow k}$ is the probability that the electron in segment $l$ comes from segment $k$ at the MB junction, $s_{Bk}$ is the length of segment $k$. The above equation can be written as a matrix equation

$$\mathbf{I} = \mathbf{M}\cdot\mathbf{V} + \mathbf{M}\cdot\mathbf{C}\cdot\mathbf{I} \qquad (9)$$

where

$$M_{lk} = P_{l\leftarrow k} \qquad (10)$$

$$V_l = \int_{s_l^{end}-s_{Bl}}^{s_l^{end}} ds' F_1(s') e^{-\int_{s'}^{s_l^{end}} F_2(s'')ds''}$$

$$C_{lk} = \delta_{lk} e^{-\int_{s_l^{end}-s_{Bl}}^{s_l^{end}} F_2(s')ds'}.$$

The $s_l^{end}$ in the above equations mean the $s$-coordinate of the end point of segment $l$. Eq. 9 is reminiscent of the Dyson equation and can be immediately solved

$$\mathbf{I} = (\mathbb{I}-\mathbf{MC})^{-1}\mathbf{MV} \qquad (11)$$

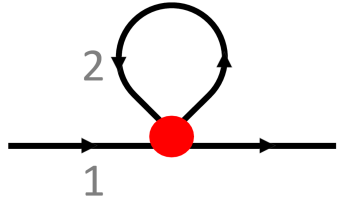

**Figure 2**. Schematics of a closed and an open orbit coupled through an MB junction.

We now use the above formalism to solve the problem of two coupled orbits illustrated in Fig. 2. We label the open orbit as 1 and the closed one as 2, both being electron-like, and with the magnetic field pointing out of the paper. At the MB junction we have

$$\mathbf{M} = \begin{pmatrix} Q & P \\ P & Q \end{pmatrix} \qquad (12)$$

where $Q \equiv 1-P$. $P=1$ means complete breakdown and $P=0$ means no breakdown. Eq. 11 then gives

$$\begin{pmatrix} I_1 \\ I_2 \end{pmatrix} = \frac{1}{(1-QC_1)(1-QC_2)-P^2 C_1 C_2}\begin{pmatrix} (1-QC_2)(QV_1+PV_2)+PC_2(PV_1+QV_2) \\ PC_1(QV_1+PV_2)+(1-QC_1)(PV_1+QV_2) \end{pmatrix} \qquad (13)$$

where $C_{1,2}$ are the diagonal elements of the $C$ matrix. One can readily check that the above solution reduces to the known results based on Eq. 3 for the special cases of $P = 0$ (no MB) and $P = 1$ (complete MB), for which we give the results of $I_{1,2}$ to be used below:

$$(I_1, I_2) = \begin{cases} \left(\frac{V_1}{1-C_1}, \frac{V_2}{1-C_2}\right) & P = 0 \\ \left(\frac{V_2+C_2 V_1}{1-C_1 C_2}, \frac{V_1+C_1 V_2}{1-C_1 C_2}\right) & P = 1 \end{cases} \quad (14)$$

When the MB junction is replaced by an effective one incorporating a small network of coherent paths, $P$ generally acquires an oscillatory part versus $H^{-1}$ [29], which can be due to either Landau quantization or quantum interference. Below we show that for the present toy model, an oscillation in $P$ leads to oscillations of $\sigma_{xx}$ and $\sigma_{yy}$ whose phases are different by $\pi$.

We first give a qualitative discussion by comparing the cases of $P = 0$ and $P = 1$ for $\sigma_{xx}$ and $\sigma_{yy}$ separately.

(1) $\sigma_{yy}$ ($\mathbf{E} \perp \mathbf{Q}$)

We consider the high-field regime and assume the mean-free path $l_{\text{mfp}} = v_\perp \tau$ to be constant for simplicity. Then

$$C_{1,2} \approx 1 - 2\pi(\omega_{c1,2}\tau)^{-1} \quad (15)$$
$$V_{1,2} \approx \int_{0_{1,2}}^{0_{1,2}+s_{B1,2}} ds' F_1(s') \left[1 - \frac{(s_{B1,2}+0_{1,2}-s')\ell^2}{l_{\text{mfp}}}\right]$$

where $\omega_{c1,2} \equiv \frac{2\pi v_\perp}{\ell^2 s_{B1,2}}$. For $V_1$ it is sufficient to keep the lowest order term since $\mathbf{v} \cdot \mathbf{E} \neq 0$ on the open orbit and its integral over the open orbit does not vanish. For $V_2$ the lowest order term vanishes [30].

For $P = 0$, the two orbits are decoupled, and for $\sigma_{yy}$ up to $O(H^{-1})$ we only need to consider the contribution due to the open orbit. Using Eq. 14 one can get

$$g_1(s) \approx g_1^{(0)}(s) = \frac{\omega_{c1}\tau}{2\pi} V_1 = \frac{e\tau}{s_{B1}} \oint_1 ds' \frac{\partial f_0}{\partial \epsilon} \mathbf{v} \cdot \mathbf{E}. \quad (16)$$

where the subscript 1 means the argument $s$ of $g_1(s)$ belongs to the open orbit, $g_1^{(0)}(s)$ means the contribution to $g_1(s)$ that is $O(H^0)$.

On the other hand, when $P = 1$, Eq. 14 suggests that

$$I_1 \approx I_2 \approx \frac{V_1+V_2}{1-C_1 C_2} \quad (17)$$

As a result

$$g_1^{(0)}(s) = g_2^{(0)}(s) = \frac{(2\pi)^{-1}}{(\omega_{c1}\tau)^{-1}+(\omega_{c2}\tau)^{-1}} V_1 = \frac{e\tau}{s_{B1}+s_{B2}} \oint_1 ds' \frac{\partial f_0}{\partial \epsilon} \mathbf{v} \cdot \mathbf{E} \quad (18)$$

Therefore

$$\frac{\sigma_{yy}(P=1)-\sigma_{yy}(P=0)}{\sigma_{yy}(P=0)} = -\frac{s_{B2}}{s_{B1}+s_{B2}} \quad (19)$$

which is always negative. Therefore for the open orbit, a finite MB probability always suppresses the conductivity perpendicular to the open orbit because the closed orbit simply increases the effective dissipation when electrons or holes from the open orbit tunnel into it, but does not contribute to current generation in the high-field limit since the carriers excited by the electric field on the closed orbit cancel out.

(2) $\sigma_{xx}$ (**E** ∥ **Q**)

In this case $V_1 = 0$ since $\mathbf{v} \cdot \mathbf{E} = 0$. Therefore, when $P = 0$ only the closed orbit contributes to $\sigma_{xx} \sim H^{-2}$. When $P = 1$, the open orbit contributes to $\sigma_{xx}$ only through $I_2 = \frac{C_1 V_2}{1 - C_1 C_2}$. As a result, the net change of $g_2(s)$ is

$$\delta g_2(s) = e^{-\int_{0_2}^{s} F_2(s')ds'} \left(-\frac{V_2}{1-C_2} + \frac{C_1 V_2}{1-C_1 C_2}\right) \tag{20}$$

However, $V_2$ in general has a finite $O(H^{-1})$ term. Assuming the closed orbit to be circular, we have

$$V_2 = eE\ell^2 \frac{\partial f_0}{\partial \epsilon} \int_{\phi_0}^{\phi_0+2\pi} d\phi \frac{s_{B2}}{2\pi} \cos\phi \, e^{-\frac{\phi_0+2\pi-\phi}{\omega_{c2}\tau}} \tag{21}$$

$$\approx eE\ell^2 \frac{\partial f_0}{\partial \epsilon} \frac{s_{B2}}{\omega_{c2}\tau} \sin\phi_0$$

where $\phi_0$ is the angular coordinate of the MB junction on orbit 2. Then one can obtain

$$\delta g_2(s) \approx e^{-\frac{\phi_s-\phi_0}{\omega_{c2}\tau}} eEl_{\text{mfp}} \frac{\partial f_0}{\partial \epsilon} (\omega_{c2}\tau)^{-1} \sin\phi_0 \left(-\frac{s_{B1}}{s_{B1}+s_{B2}}\right) \tag{22}$$

To get the contribution from $\delta g_2(s)$ to the current, we use Eq. 4 and for simplicity consider a 2D system:

$$\mathbf{j} = -\int \frac{dsd\epsilon}{(2\pi)^2} \frac{e}{\hbar} \hat{v}(s) g(s,\epsilon) \tag{23}$$

We also consider zero temperature so that $\partial_\epsilon f_0 = -\delta(\epsilon - \epsilon_F)$. To perform the $s$ integral one needs to pay attention to how to convert $s$ defined for the two orbits joined together into angular variable $\phi$. The resulting $\delta\sigma_{xx}$ is

$$\delta\sigma_{xx} = \frac{e^2}{h} \frac{1}{E} \int_{\phi_0}^{\phi_0+2\pi} \frac{s_{B2}}{(2\pi)^2} \cos(\phi_s) \delta g_2(\phi_s) d\phi_s \tag{24}$$

$$\approx \frac{e^2}{h} \frac{s_{B2} l_{\text{mfp}}}{2\pi} (\omega_{c2}\tau)^{-2} \left(\frac{s_{B1}}{s_{B1}+s_{B2}}\right) \sin^2\phi_0$$

which is of the same order $O(H^{-2})$ as $\sigma_{xx}(P=0)$ and is, more importantly, non-negative. Therefore as long as $\phi_0 \neq 0$, i.e. the open orbit is not exactly bisecting the closed one, a finite tunneling probability $P$ can enhance $\sigma_{xx}$.

The reason for the non-negative $\delta\sigma_{xx}$ can be understood in the following way. Different from the open orbit, for which a higher dissipation always suppresses conductivity, the conductivity for the closed orbit increases when $\omega_c \tau$ becomes smaller in the high-field limit. Since the only effect of the open orbit for $\sigma_{xx}$ is to increase dissipation, in the high-field limit coupling to it can make the conductivity for the closed orbit larger.

We finally consider the Hall conductivities. For simplicity we only discuss $\sigma_{yx}$ below since it can be shown for the present model that $\sigma_{xy} = -\sigma_{yx}$. In this case $V_1$ vanishes identically. But since the velocity along $y$ for the open orbit is nonzero, it can contribute to $\sigma_{yx}$ when $P = 1$. Namely, $g_1(s)$ becomes nonzero when $P = 1$. Using Eq. 21, we obtain:

$$\delta g_1(s) = e^{-\frac{s\ell^2}{l_{\text{mfp}}}} \frac{V_2}{1-C_1 C_2} \tag{25}$$

$$\approx e^{-\frac{s\ell^2}{l_{\text{mfp}}}} \frac{s_{B2}}{s_{B1}+s_{B2}} l_{\text{mfp}} (\omega_{c2}\tau)^{-1} \sin\phi_0$$

On the other hand, $\delta g_2(s)$ is given by Eq. 22 and is already $O(H^{-1})$, which will lead to an $O(H^{-2})$ contribution to the current after integration over the closed orbit. Therefore we only need to consider the current or $\delta\sigma_{yx}$ due to $\delta g_1$:

$$\delta\sigma_{yx} = -\frac{e^2}{h}\frac{s_{B2}}{s_{B1}+s_{B2}}l_{\text{mfp}}(\omega_{c2}\tau)^{-1}\sin\phi_0 \int_0^{s_{B1}} \frac{ds}{2\pi} e^{-\frac{s\ell^2}{l_{\text{mfp}}}} \quad (26)$$
$$\approx -\frac{e^2}{h}\frac{(s_{B2})^2 s_{B1}}{(2\pi)^2(s_{B1}+s_{B2})}\ell^2\sin\phi_0$$

which can be shown to be identical to $-\delta\sigma_{xy}$. The sign of $\delta\sigma_{xy}$ or $\delta\sigma_{yx}$ is therefore non-universal as it depends on the sign of $\sin\phi_0$ in addition to the carrier types of the two orbits. In the present model, when $\sin\phi_0 < 0$ as in Fig. 2, $\delta\sigma_{xy}$ and $\delta\sigma_{yy}$ are in phase and are out of phase with $\delta\sigma_{xx}$.

Numerical results in Fig. 6b of the main text were obtained using Eqs. 6 and 13 by considering a 2D system and zero temperature, assuming $P = P_0 + P_1\cos(\ell^2/\ell_0^2)$, with the other parameter values given in the table below. The conductivities thus have the units of $e^2/h$ and are plotted against $\ell^2/l_{\text{mfp}}^2$.

**Table 1.** Parameter values used for getting the numerical results in Fig. 6 of the main text.

| | |
|---|---|
| $\ell_0^2$ | 0.1 |
| $P_0$ | 0.6 |
| $P_1$ | 0.1 |
| $l_{\text{mfp}}$ | 1 (length unit) |
| $s_{B1}$ | 0.2 |
| $s_{B2}$ | 1.0 |
| $\phi_0$ | $-\pi/2$ |